\documentclass[9pt,twocolumn]{article}
  \pdfoutput=1\relax                   
  \usepackage{graphicx}                
  
\graphicspath{{figures/}{pictures/}{images/}{./}} 

\usepackage{microtype}                 
\PassOptionsToPackage{warn}{textcomp}  
\usepackage{textcomp}                  
\usepackage{mathptmx}                  
\usepackage{times}                     
\usepackage{cite}                      
\usepackage{tabu}                      
\usepackage{booktabs}                  
\usepackage{amsmath}
\usepackage{amssymb} 

\usepackage{float}
\usepackage{color}
\usepackage{listings}
\definecolor{mygreen}{RGB}{0, 153, 51}

\newcommand{\oursystem}{DimReader}
\usepackage{geometry}

\title{\oursystem{}: Axis lines that explain non-linear projections}

\author{Rebecca Faust, David Glickenstein and Carlos Scheidegger}
\date{\includegraphics[width=\linewidth]{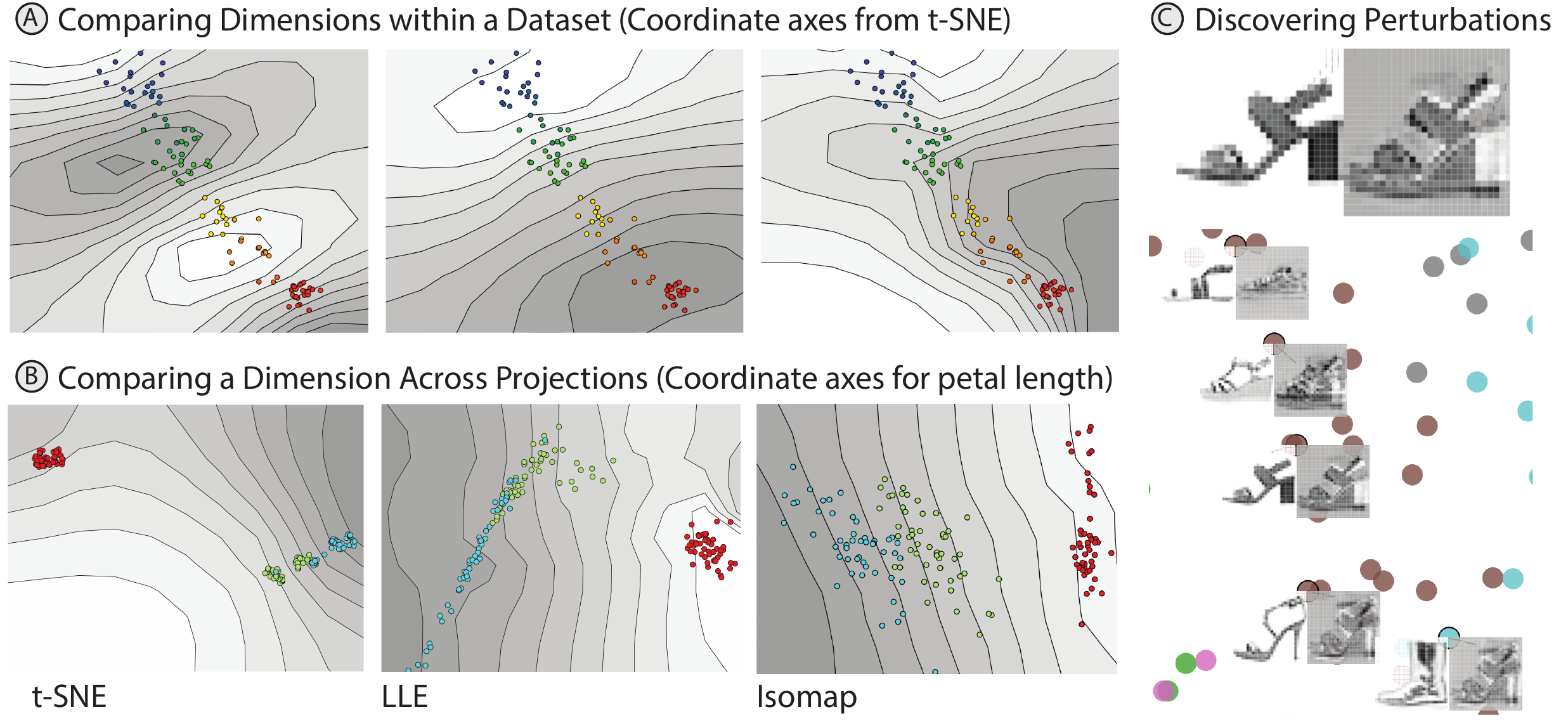}}

\newcommand{\cs}[1]{}
\newcommand{\rf}[1]{}
\newcommand{\new}[1]{{\color{black}#1}}

\begin{document}

    \twocolumn[
  \begin{@twocolumnfalse}
    \maketitle
    \vspace{-5mm}
     Figure 1: \oursystem\ explains non-linear dimensionality reduction methods by illustrating the effects of user-designed perturbations of the input dataset. It provides an answer to the question ``if the input data had been slightly different in a particular way, how would the plot have changed?''. In the case of traditional scatterplots, it recovers exactly the axis lines being displayed. In the case of non-linear methods, \oursystem\ recovers \emph{generalized axes}, which indicate how dimensions of interest behave.  Examples of these axes are shown in (A) for the x, y, and z dimensions of the S Curve (an S shaped 3 dimensional manifold). These axes also allow for the comparison of different projection methods. This is exemplified in (B), where the petal length axis of the iris dataset is shown for three projections.  Petal length is well behaved in t-SNE but not in the other projections .  We also provide a technique for discovering good perturbations of the input (perturbations that change the projection the most). The top of (C) shows an example of a discovered perturbation. In context, shown at the bottom of (C), this perturbation shows us that t-SNE is sensitive to flat shoes v.s. heels.  The perturbation wants to change the original image from a heel to a flat by filling in the arch.
    \vspace{-2mm}
  \paragraph{Abstract-}Non-linear dimensionality reduction (NDR) methods such as LLE and t-SNE are popular with visualization researchers and experienced data analysts, but present serious problems of interpretation. In this paper, we present \oursystem{}, a technique that recovers readable axes from such techniques. \oursystem\ is based on analyzing infinitesimal perturbations of the dataset with respect to variables of interest.  The perturbations define exactly how we want to change each point in the original dataset and we measure the effect that these changes have on the projection. The recovered axes are in direct analogy with the axis lines (grid lines) of traditional scatterplots.  We also present methods for discovering perturbations on the input data that change the projection the most.  The calculation of the perturbations is efficient and easily integrated into programs written in modern programming languages. We present results of \oursystem\ on a variety of NDR methods and datasets both synthetic and real-life, and show how it can be used to compare different NDR methods. 
Finally, we discuss limitations of our proposal and situations where further research is needed.
\vspace{5mm}
  \end{@twocolumnfalse}
]

\section{Introduction}
One of the central promises of data visualization is that its
techniques will help users and analysts make sense of large,
complicated datasets.  Data visualization, and specifically techniques
in dimensionality reduction, are routinely used in practice during
exploratory data analysis of challenging datasets.

Classical linear methods such as Principal Components Analysis have
existed for more than a century, but recent advances from
\emph{non-linear} methods that started with Tenenbaum et al's
Isomap~\cite{tenenbaum2000global} have revolutionized the practice of
dimensionality reduction. The potential to understand high dimensional
data via low-dimensional representations is clearly attractive. But
just what, exactly, are these non-linear dimensionality reduction
(NDR) methods showing? This is the fundamental question that drives
the work we report here.  Data scientists and analysts use NDR's in an
attempt to create a nice 2-dimensional representations for their data
in hopes of learning some of the underlying structure of the
data. The NDR's often result in nice pictures but give no indication
\emph{why} the NDR placed things the way it did and no context to
the input.
   
\begin{figure*}
  \centering
    \includegraphics[width=.9\linewidth]{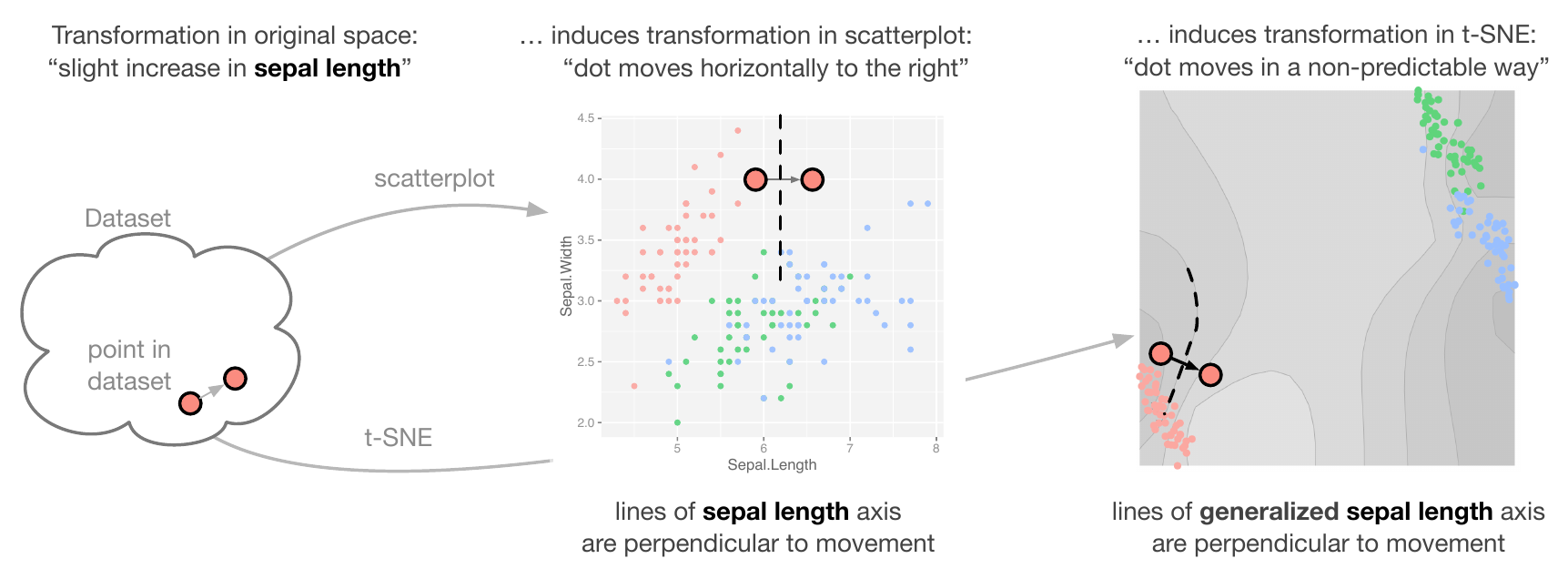}
  \caption{In traditional scatterplots, the grid lines (or axes lines) exist to explain what the plot is showing. Equivalently, they capture \emph{infinitesimal perturbations} of the dataset in specific directions, because they are always perpendicular to the directions of movement. \oursystem\ extends the same principle to non-linear dimensionality reduction (NDR) methods, and recovers \emph{generalized axis lines}, which help explain NDR methods in terms of interpretable data transformations.\label{fig:explainer}}
\end{figure*}

Consider van der Maaten and Hinton's t-SNE, arguably the most powerful
and currently most popular method for
NDR~\cite{maaten2008visualizing}. Although practical experience
attests to t-SNE's power to uncover cluster relationships in very
challenging datasets, its sensitivity to the hyper-parameters is
remarkable~\cite{wattenberg2016how}. If small changes in parameter
settings produce plots that are fundamentally different, we must ask
ourselves: are some results generated by NDR methods just bad? Do
different parameter settings show different features of the data? More
importantly, how do we even answer these questions?
  
In this work, we design data transformations, which induce
transformations on the visualization itself, elucidating
the behavior of the NDR method (this is the perspective introduced by Kindlmann and
Scheidegger's algebraic design process~\cite{Kindlmann2014algebraic}).
Specifically, we use infinitesimal perturbations ---
small changes of the data in its original space --- to produce
infinitesimal changes of the visualization. We then show how these
visualization changes can be interpreted as producing \emph{effective,
  non-linear axis legends}. In this way, our non-linear axes \emph{explain} the NDR plot in the same
way that axis legends explain the positional encoding in scatterplots.
As a result, analysts can understand
and evaluate dimensionality reduction plots similarly to how they
evaluate \emph{linear} methods. In fact, we show in
Section~\ref{sec:technique} that our methods exactly recovers
the gridlines of typical scatterplots. 
\oursystem\ is quite general, and can be applied to many different
NDR techniques, only requiring access to the source code of its
implementation. Specifically, we use a method known as \emph{automatic
  differentiation} to produce the necessary gradients for calculating the infinitesimal changes of the visualization~\cite{griewank2008evaluating}. An overview of the process is given in Figure~\ref{fig:explainer}.

In summary, our contributions are:

\begin{itemize}
  \item A general framework to explain plots generated by non-linear dimensionality reduction, using infinitesimal perturbations
   \item A practical implementation of the framework using automatic differentiation
  \item A method for discovering good perturbations for a given dataset, useful when the input lacks easily interpretable dimensions (and hence, lacks easily-defined perturbations)
  \item An experimental study of the effectiveness and efficiency of \oursystem\ using three well-known NDR methods: Isomap~\cite{tenenbaum2000global}, LLE~\cite{roweis2000nonlinear}, and t-SNE~\cite{maaten2008visualizing}.
\end{itemize}

\begin{figure*}
  \centering
  \includegraphics[width=.8\linewidth]{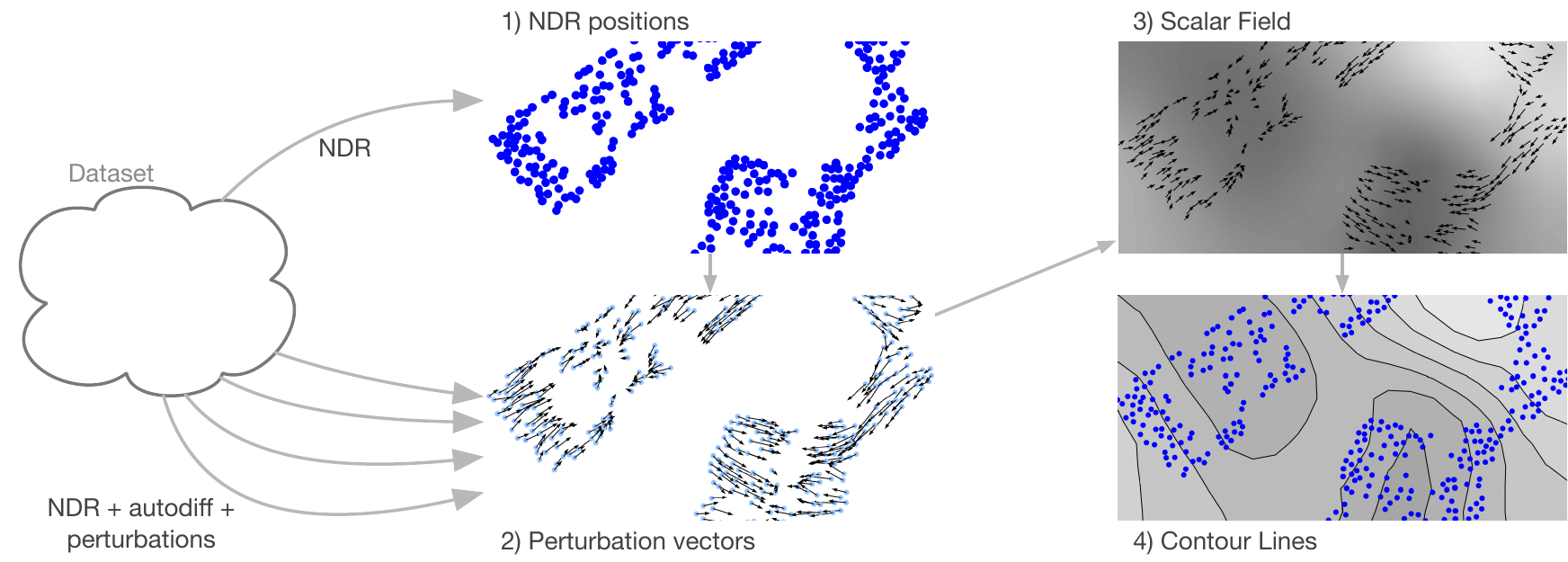}
  \caption{An overview of \oursystem. For a given NDR method, we 1) compute its position using the original implementation; 2) compute perturbation directions for the input points with the transformed version of the implementation which uses dual numbers (We discuss how to choose appropriate perturbations in Section~\ref{sec:whichperturbation}); 3) compute the scalar field whose gradient best matches the perturbation vectors in a least-squares sense; and finally 4) compute its isocontours. Section~\ref{sec:technique} explains these steps in detail.\label{fig:method-explainer}}
\end{figure*}

\section{Related Work}

Projection methods have received a considerable amount of attention in information visualization. In this section, we review the work that is most directly related to our research, but cannot hope to cover the entirety of the field. For a comprehensive view on multidimensional scaling and dimensionality reduction, we recommend Born and Gr\"{o}nen's textbook~\cite{borg2005modern}, and Fodor's survey~\cite{fodor2002survey}.

\paragraph*{Projection methods in information visualization} The observation that pairwise similarities (or distances) can be converted into low-dimensional representations by a mathematical formulation comes from Torgerson and his now-classical theory of multidimensional scaling~\cite{torgerson1965multidimensional}. In information visualization, force-directed methods have long been used as a dimensionality reduction technique, from fully-automatic methods~\cite{chalmers1996linear, morrison2002hybrid, ingram2009glimmer}, to methods which take some amount of interaction, either through placement of exemplar points~\cite{paulovich2008least, joia2011local, endert2011observation} or through direct interaction with projection parameters~\cite{jeong2009ipca}. Although interactive methods offer a better hope for understandability because the perturbation analysis we discuss can happen ``in the analyst's head'' during interaction, we argue that the visual encoding these techniques provide can still be unclear. The technique we propose here can be applied to essentially all of the methods above, and offers an attractive complement to both automated and interactive projection methods.

\paragraph*{Perturbation Analysis for data science} The idea of understanding a system by examining its behavior under perturbations is well-established in the engineering and statistics literature. In the 1970's, Cook introduced the notion we now know as Cook's distance~\cite{cook1977detection}, which measures the influence of a point on the parameters of linear regression models. In the context of visualization, Bergner et al. point to sensitivity analysis as one of the requirements in understanding computer simulations~\cite{bergner2013paraglide}. In this paper, we use perturbation analysis as a central tool to recover readable axes from NDR methods, in a sense incorporating sensitivity analyses into familiar visual metaphors. 

\paragraph*{Automatic Differentiation} Perturbation analysis is clearly an important tool for understanding systems, but the issue of how to implement it in existing computer systems is crucial. Automatic differentiation (which we explain in detail in Section~\ref{sec:technique}) provides a way to compute derivatives of arbitrary functions in a computer program, provided access to the source code (or similar structural information about the computation) is available~\cite{griewank2008evaluating}. To the best of our knowledge, the most mature software library employing automatic differentiation is Ceres, written in C++ and employing template metaprogramming~\cite{ceres-solver}. \oursystem\ is implemented in Python for simplicity and terseness, but could easily be redesigned in C++.

\paragraph*{Guidance and validation of projection results} One of the issues with NDR is that it's hard to know what a plot is actually showing\cite{vellido2012making}. This has resulted in a variety of papers which offer guidance and design principles on how to interpret projections, based on a combination of real-world experience, synthetic examples, and theoretical arguments~\cite{buja2002visualization, sedlmair2012dimensionality, sedlmair2013empirical, sacha2017visual,lee2008quality}. This work is essential to the current practice, we argue, because current NDR methods do not offer explanations of their own results --- there are much fewer research papers offering guidance for understanding and interpreting traditional scatterplots. As we show in Section~\ref{sec:implementation}, our technique provides a way for a projection method to explain itself. Although analyst guidance and validation will always be a part of a well-designed analysis infrastructure, our technique could mitigate some of the problems that have been observed in deployed systems, where projection methods are ultimately discarded because of readability issues~\cite{brehmer2014overview, ingram2017personal}.
\rf{Added reference to paper on measures to asses quality of NDR in above list about papers that offer guidance, not sure if that is correct}

\paragraph*{Augmented visual representations} There is another avenue of attack on the readability problem of NDR methods. Often, researchers will \emph{augment} the results of the projections with visual diagnostics that pinpoint potential problems. 
Seifert et al. augment the projection by showing how the projection's stress (roughly the discrepancy between source-space distances and target-space distances) varies spatially in the NDR plot~\cite{seifert2010stress}.   \new{ In ~\cite{aupetit2007visualizing} and  ~\cite{lespinats2011checkviz}, the projection is augmented to show uncertainty measures and distortions in NDR's respectively.  Cutura et. al's VisCoDeR allows users to compare and explore different dimensionality reductions by augmenting the projection to allow users to explore how dimensions are mapped in the dimensionality reduction results as well as the high-dimensional proximity of projected points to a selected point in the projection ~\cite{cuturaviscoder}.\rf{This citation is incomplete (I just copied from google scholar)} } Stahnke and co-authors described methods to \emph{probe} a projection, through carefully designed user interactions and custom visual encodings~\cite{stahnke2016probing}. Our method for extracting effective axes can be seen as a way to allow \emph{any} NDR method to augment itself with metaphors that have a well-defined analogy in the linear case, as can be seen in Section~\ref{sec:implementation}, and Figure~\ref{fig:iris-all} specifically. In Section~\ref{sec:synthetic}, we provide a more direct comparison to some of the methods used in Stahnke et al.'s work. 

\paragraph*{Explainable visualizations} Every plot assumes an audience that can read it, and visualization literacy remains an active area of research~\cite{boy2014principled}. Often, novel metaphors are necessary because of the data or task complexity~\cite{brehmer2013multi}. We argue that generalizing well-established techniques such as axis legends to NDR can help explain those techniques. Gleicher's Explainers take user interaction to design specific projections for input data~\cite{gleicher2013explainers}. In contrast, our technique extracts axes inherent in the non-linear projections.  Coimbra et al. explain projections through enhanced biplots~\cite{coimbra2016explaining}.  Similar to our technique, they show axes for the dimensions of the input data in the low dimensional plot.
Because of the non-linearity of these dimensionality reductions, the biplot axis will change based on the projected position of the sampled point whereas our technique captures the axis lines for the entire projection. A similar approach is proposed by Cavallo and Demiralp in Prolines, a technique for interacting with data points in both low- and high-dimensional spaces~\cite{cavallo2018visual}. Prolines allow efficient, direct manipulation of the output points, but require access to efficient forward and backward projection, limits its applicability. Flow-based scatterplots ~\cite{chan2010flow} and Generalized Sensitivity Scatterplots ~\cite{chan2013generalized} show the sensitivity of a dimension in a scatterplot with respect to other dimensions in the dataset.  Similar to our technique, these methods use derivatives to determine the sensitivities.  Our technique differs by showing sensitivity of the projection with respect to the original data rather than sensitivity between dimensions in the original data. The data context map from Cheng et. al. provides a way to simultaneously look at clusters of data points and the location of the most dominant values of each attribute with the assumption that the attribute values always decrease as points move farther away from it. ~\cite{cheng2016data}. Our technique differs by showing the behavior of a dimension throughout the entire projection, not just the location of the most dominant value.



\section{Technique\label{sec:technique}}
Our technique is broken into two parts: (1) explaining an NDR method using a known perturbation (\oursystem\ ) and (2) searching for good perturbations when there is no known perturbation (after which \oursystem\ can be applied).  

In principle, all that \oursystem\ requires is the ability to
compute derivatives of the projection coordinates with respect to each of the input
points. For extremely simple techniques (such as scatterplots and
other fixed linear projections), these derivatives can easily be
evaluated in closed form. However, more sophisticated methods such as
Isomap, LLE, and t-SNE involve long computation chains, for
which the evaluation of the derivative would introduce significant
development overhead. Instead of trying to solve them in closed form,
we take central advantage of automatic differentiation.~\cite{griewank2008evaluating}. 
As we describe next, automatic differentiation allows us to calculate 
the derivative of a projection with minimal implementation effort.

\subsection{Automatic Differentiation}

In this paper, we use a particular form of automatic differentiation
known as forward-mode automatic differentiation. In what follows, we will
refer to it as ``autodiff''.

In forward-mode autodiff, the
program's derivative with respect to a specified variable is computed
alongside the function value, by using an \emph{extended number
system}.  In this system, we replace numbers in the program with dual numbers that have the form 
$x = (a,b)$ where $a$ holds the original value of the number and $b$ carries the 
derivative of $x$  with respect to our variable of
interest.  When we initialize a variable $y$, we set $b$ to one if that is the variable we want to differentiate 
with respect to (since $dy/dy = 1$) and zero otherwise.  When the projection is run with dual numbers in place of 
regular numbers, in addition to calculating the projected points, it calculates their derivatives through applications of the chain rule and derivative rules (product rule, quotient rule, etc.).


Note that autodiff is always performed at a
specific value, and with respect to a specific variable.  It produces
two numbers as a result: the function value and the partial derivative
with respect to the chosen variable. This has two important
consequences for our design.  First, we need to decide over exactly
which variables we will take derivatives. Second, we need to execute
the program many times in order to evaluate many different
derivatives. This will become important in
Section~\ref{sec:performance}.

\subsection{DimReader Process}

\subsubsection{Overview of the process}

To apply \oursystem\ to an NDR method, there are four steps. Each of these steps is discussed in a subsection below.

\begin{itemize}
  \item A user chooses a perturbation of interest, which defines an infinitesimal change for each data point (possibly in different directions).
  \item The NDR method is executed many times using dual numbers, from which we obtain the perturbation vectors, one for each input point.
     \item From the perturbation vectors, a scalar field whose gradient matches the perturbation vectors is computed.
  \item The isolines of this scalar field, which are perpendicular to the gradient, are extracted using Marching Squares. They form the effective axes.
\end{itemize}

\begin{figure}
  \centering
\begin{lstlisting}[language=Python]
# Basic method, O(numPoints) runs
for i in range(0, numPoints):
    points = copy(inputPoints)  
    points[i] = perturb(points[i], perturbation)
    projection = project(points) # project uses autodiff
    dx, dy = projection.derivative[i]
    projectionVectors[i] = vector(dx, dy)
return projectionVectors

# Improved method, O(log(numPoints)) runs
counts = zero_array(numPoints)
projectionVectors = zero_matrix(numPoints, 2)
while any(counts < 1):
    points = copy(inputPoints)  
    for i in range(numPoints):
        if random() < 0.5: # perturb each point with probability 0.5
            perturbed[i] = true
            points[i] = perturb(inputPoints[i], perturbation)
    projection = project(points) # project uses autodiff
    for i in range(numPoints):
        if perturbed[i]: # only store vectors of perturbed points
            dx, dy = projection.derivative[i]
            projectionVectors[i] += vector(dx, dy)
            counts[i] += 1
for i in range(numPoints): # average all perturbations performed
    projectionVectors[i] /= counts[i]
return projectionVectors
%\end{minted}
\end{lstlisting}
\caption{Although a basic implementation of \oursystem\ is easy to understand (top), it only extracts one perturbation vector at a time. A more efficient implementation (bottom) extracts half of the perturbation vectors from the input at once. To remove possible correlations between the outputs, we choose which points to include at random, and iterate until all points have been included. The expected time in this case is logarithmic on the size of the input point dataset.\label{fig:many-at-a-time}\label{fig:extract-perturbation}}
\end{figure}

\subsubsection{Choosing which perturbation to use\label{sec:whichperturbation}}

The first step of our method involves a choice of the perturbation of the dataset. \new{A \textbf{perturbation} is a small change to a specific dimension (or set of dimensions) for each data point in the original, high-dimensional space.  Thus, the choice of perturbation corresponds}, effectively, to an analyst answering the following question: ``if each data point were slightly different in this specific way, what would happen to the visualization?'' In order to recover different features of the NDR method and its effect on the dataset of interest, different perturbations can be designed. In the following, we discuss choosing a perturbation in a dataset with interpretable dimensions.  We discuss discovering perturbations for other datasets in Section ~\ref{sec:searchingPerts}.  \new{In automatic differentiation, perturbations are represented by the derivative part of the dual number for the original data points.  The perturbation of a data point with $d$ dimensions has the form of a unit vector with $d$ elements where the value of each element specifies how much the corresponding dimension is perturbed relative to the rest of the dimensions.}

\paragraph*{Datasets with interpretable dimensions} Some datasets have interpretable columns. Take the iris dataset, for example, which is used in Figure~\ref{fig:explainer}. In that case, a perturbation that changes each of the input points in the direction of a given dimension will reconstruct, for an NDR method, curved axes lines that correspond, roughly to the linear grid lines in scatterplots. Concretely speaking, we evaluate each input point $p_i$ as $(p_i ,(0, \cdots, 0, 1, 0, \cdots, 0))$, where the value $1$ is positioned at the dimension of interest.

\subsubsection{Extracting derivatives from NDR methods}


In this section, we describe two techniques used in \oursystem\ to extract the \textbf{perturbation vectors} \new{,the changes to the projected coordinates resulting from a perturbation on the input,} for a given projection. The first technique is simple, straightforward, and provides a good intuition for the overall strategy. Unfortunately, this technique requires as many executions of the NDR method as there are input points in the dataset, which often means the overall performance can suffer. The second technique, on the other hand, only requires as many runs as the \emph{logarithm} of the number of input points. We give pseudo-code for the two approaches in Figure~\ref{fig:extract-perturbation}.

\oursystem\ needs access to the source code for the NDR method at this step so the method can be executed with dual numbers. In principle, the source code can be executed without any modifications aside from converting the input points into dual numbers. In practice, some issues arise because of efficiency concerns and library limitations. We discuss these issues at length in Section~\ref{sec:implementation}.




\begin{figure}
 \centering
  \includegraphics[width=\linewidth]{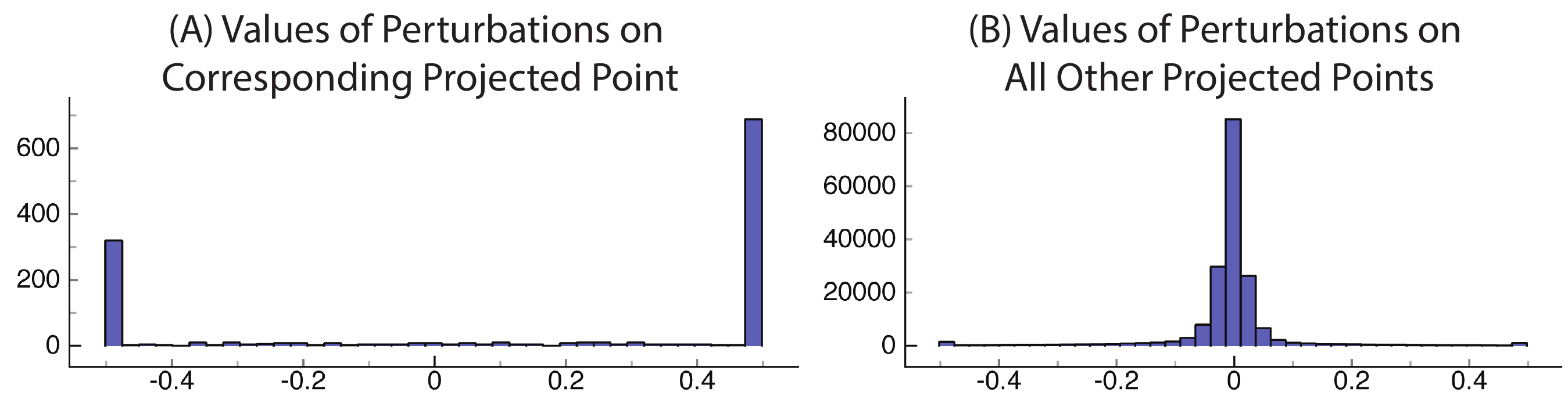}
  \caption{(A) shows the effect sizes of perturbing point $p_i$ on its corresponding projected point, $v_i$. DimReader uses these values in its computations: note their large magnitude (values outside of the displayed range are clamped at -.5 and .5). (B) shows the effects sizes of perturbing point $p_i$ on all projected points aside from $v_i$. DimReader assumes these values are zero and discards them: note their small magnitude.
  \label{fig:hists}
  }
\end{figure}

\paragraph{Perturbing one point at a time}
After a perturbation is chosen, the NDR technique is executed with automatic differentiation for every point in the dataset.  On execution $i$, the point $p_i$ is perturbed (that is, we replace $p_i$ with $(p_i, \bar{p_i})$ where $\bar{p_i}$ is the specified perturbation of $p_i$).  The NDR technique will return the projection coordinates, $v$, for all points, along with the derivative of the projection coordinates with respect to the perturbation of $p_i$, $\frac{dv}{dp_i}$.  We use the derivative of each coordinate in the reduced point $v_i$ as the vector that describes the change in the coordinate, and discard the rest of the information of the run.  The pseudocode for this is given on the top half of Figure~\ref{fig:extract-perturbation}.

\paragraph{Perturbing many points at a time}
The method described above is inefficient, requiring $O(n)$ evaluations of the NDR method. A naive attempt to optimize the method would evaluate the projection derivatives with respect to all of the points (and hence all of the per-point perturbations) at once, and only run the autodiff version of the code once. Unfortunately, this does not work for many perturbations, because most DR methods are \emph{invariant to dataset translations}. The perturbation of only one input point at a time offers interesting insight into the NDR method, but if we move all of the points at once in the same direction, NDR methods such as Isomap, LLE, and t-SNE will produce exactly the same projection (the perturbation vectors will be all zeros).

We solve this problem by adding a small amount of randomization. Instead of perturbing one point at a time, we can choose half of the points at random to perturb, while the other half does not change. We then store the projection vectors for the points we chose to perturb, and repeat the process until we have actually perturbed all of the input points. After each round, we expect to halve the number of unperturbed points, which gives an expected number of repeated runs which is logarithmic on the number of input points. The pseudocode for this is given on the bottom half of Figure~\ref{fig:extract-perturbation}. We found that the DimReader plots produced by perturbing many points at a time are indistinguishable from the plots produced by perturbing one point at a time.  

\paragraph{Ignoring changes in unperturbed points} In some cases, perturbing a point $p_i$ has an effect on points other than its corresponding projected point $v_i$. However, the effects on other points are small enough that we can effectively ignore them. In Figure ~\ref{fig:hists}, we show that the effect of perturbing each point $p_i$ on the other points, $v_j$ ($i\neq j$) tends to be near zero and very rarely is significant.  We show this result for the iris dataset, but have found it to be true in general for t-SNE in all datasets we checked. Intuitively, we expect a good dimensionality reduction to be robust to a small change in a single point, and thus the residual effect on the rest of the points to be insignificant. 

\subsubsection{Reconstructing the direction field}

\begin{figure}
  \centering
  \includegraphics[width=.8\linewidth]{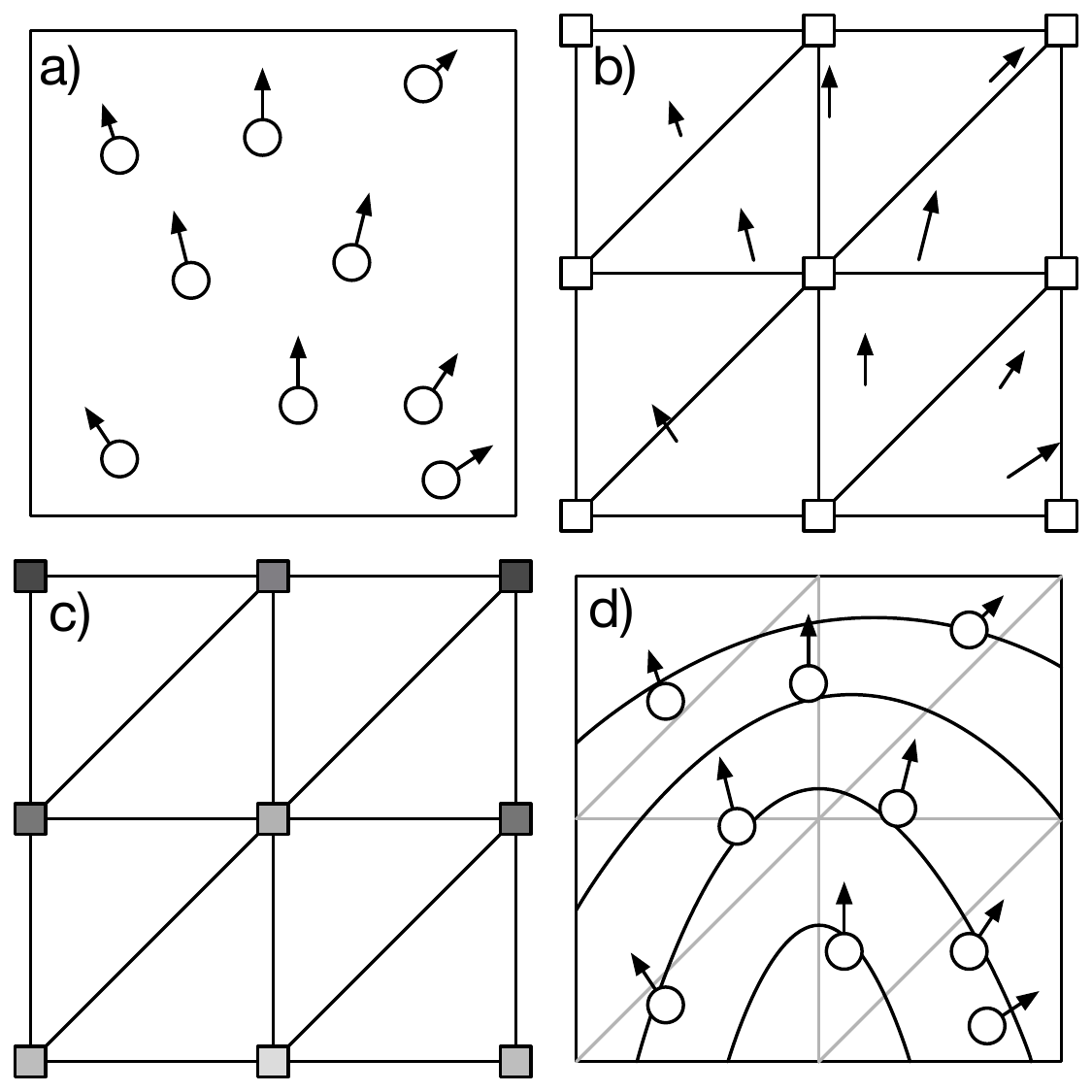}
  \caption{An illustration of the process to recover generalized axes. Given the point positions and perturbation vectors (a), we construct a triangular mesh and interpret each vector as a linear constraint on the gradient of a function (b), which gives values on each of the vertices (c). From these values, we can extract lines perpendicular to the perturbation vectors using marching squares.\label{fig:vectors-to-isocontours}}
\end{figure}

Once we have the projected points and their derivatives (that is, the perturbation vectors), we need to reconstruct the direction field, in order to extract perpendicular lines.
We achieve this by computing a scalar field whose gradient best matches the vectors. We use a simple least-squares reconstruction technique, adapted from Ferreira et al.'s vector-field clustering work~\cite{ferreira2013vector}, which we illustrate in Figure~\ref{fig:vectors-to-isocontours}. We first decompose the output plane in a rectangular grid, and split each grid square into two triangles, giving a triangular mesh of the output space. The resolution of this grid needs to be decided ahead of time, and we use a 10x10 grid in our examples for this paper. We model a scalar field on the output plane as a piecewise-linear function on the grid values. Each point and its perturbation vector is interpreted as a linear constraint on the vertices of its corresponding triangle. To find the best-fitting scalar field, we solve it in a least-squares sense, regularizing the system to ensure a unique solution~\cite{ferreira2013vector}. \rf{Reviewer 2 said more detail, we could add more about how we get the scalar values on the vertices from the triangles but it seems like we would need to add the equations and stuff which would add quite a bit of content}



     		

\begin{figure}
  \centering
   \includegraphics[width=.78\linewidth]{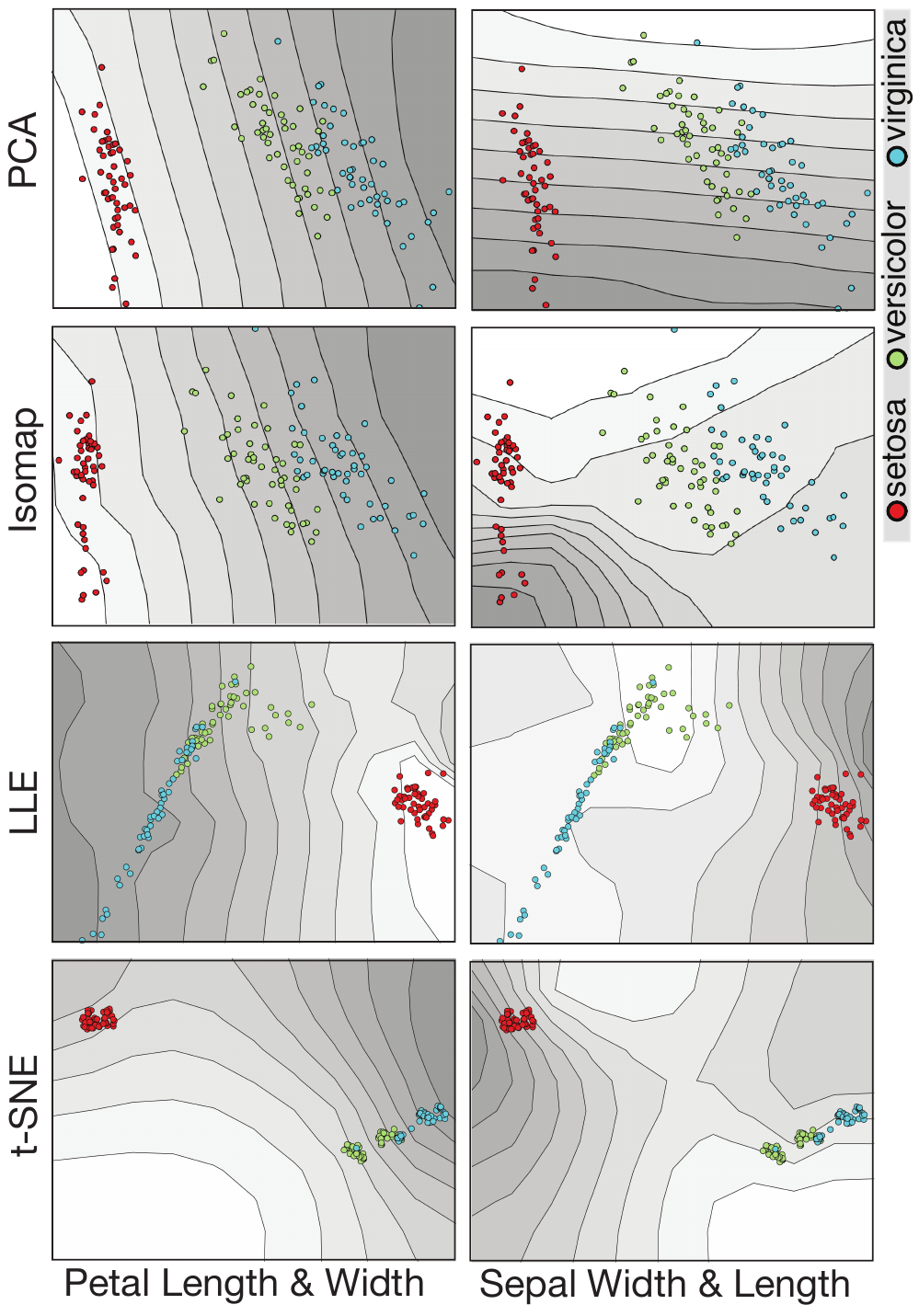}

  \caption{Extracting axes from the Iris dataset with four projections: PCA, Isomap, LLE, and t-SNE.  We only show petal length and sepal width because petal width is extremely similar to petal length and sepal length is very similar to sepal width. We discuss how to interpret these plots in Section~\ref{sec:dimReaderExpir}
   \label{fig:iris-all}}
\end{figure}
\subsubsection{Extracting perpendicular lines}

The final step is quite simple. With the scalar field expressed as values in a triangular mesh, we can use marching squares to extract isocontours~\cite{angel2003interactive}. By construction, the gradient of this scalar field matches the perturbations. Since isolines are perpendicular to a function's gradient~\cite{schey1973div}, the resulting curves will tend to be perpendicular to the perturbations. As we show in Figure~\ref{fig:explainer}, these isoline can be thought of as \emph{generalized axes lines}.

\subsection{Interpreting DimReader Plots}
\begin{figure}[t]
  \centering
    \includegraphics[width=\linewidth]{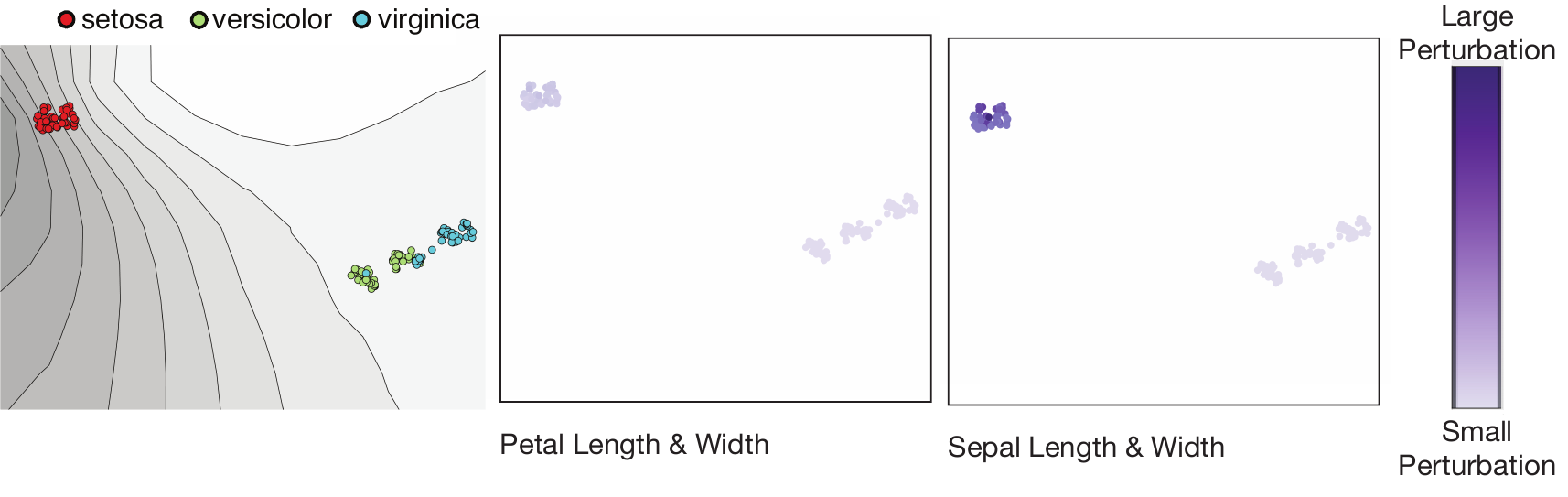}
  \caption{\label{fig:IrisPert} The best perturbation for the iris dataset.  The left plot is the DimReader plot for this perturbation.  In the two plots on the right, the color of each point shows how much the point is perturbed for the specified dimension.  We see that the sepal width and sepal length are perturbed more in the Red cluster (the Setosa cluster) than the petal dimensions which means that, for this cluster, the projection is sensitive to changes in the Sepal dimensions. The perturbations for the points in the other cluster are insignificant.  This tells us that perturbing only the Setosa points will change the projection the most.}
\end{figure}

In our plots, the projected points would move perpendicular 
to the isolines \new{nearest to them} if the input were perturbed in the specified way. 
An increase in the corresponding dimension would move the point from light to dark.
The relative density of the isocontours can be interpreted
similarly to the behavior in scalar fields. Narrowly-spaced
isocontours indicate a high sensitivity to changes in the
\emph{independent} variable, (in our case, projection coordinates).
Widely-spaced isocontours indicate \emph{low spatial sensitivity}: a
change in the projection coordinates is not expected to change the
outcome variable by much.  Curved isolines indicate that the same perturbation 
has a different effect on different points. Isolines that fan out 
(go from narrowly-spaced to widely-spaced as in Figure ~\ref{fig:IrisPert}) indicate that 
the sensitivity of the plot is changing from more sensitive on one side to less sensitive on the other.



\subsection{Discovering Good Perturbations}\label{sec:searchingPerts}
We may not always know good perturbations for a dataset, such as the MNIST digits where it is not clear what the best way to perturb each image would be. To help solve this problem, we offer two methods to recover good perturbations. We define good perturbations as perturbations on the input that change the projection the most under the given constraints.  Both of the methods require that we have a tangent map, $M$, for the projection. \new{The tangent map allows for efficient calculation of the perturbation vector, $\bar{v}$, for a given projection, without running the projection itself. Given a perturbation on the input $\bar{p}$, $M \cdot \bar{p}$ results in the perturbation vector $\bar{v}$, i.e. $\bar{v} = dv/dp$ where $v$ is the projected coordinates}.
The vector $\bar{p}$ consists of the perturbation of each input point concatenated into a single vector (end to end) \new{and the perturbation vector $\bar{v}$ consists of the perturbation vector $\bar{v_i}$ of each projected point ($v_i$) concatenated into a single vector.}  A single column of $M$ can be recovered by a perturbation vector that has a single entry of 1 and the rest zeros (i.e. perturbing a single dimension of a single point).  By doing this for each dimension of every point, the entire tangent map can be recovered.  

One observation about the tangent map is that the values we need for calculating the perturbation vectors lie in $k \times d$ blocks along the diagonal, where $k$ is the dimension of the projection (typically 2) and d is the dimension of the input data.  Because we ignore the effect a given perturbation on all other points (as discussed in ~\ref{sec:whichperturbation}), we set the rest of the matrix to zero.  We exploit the block structure of the tangent map in both of our methods for finding the best perturbation. 


 
\subsubsection{Perturb all points in the same direction} The first method for recovering the best perturbation is to find the single perturbation that when applied to all points changes the projection the most.  The formulation for this method is: 

\begin{equation*}
\underset{\bar{v}\in\mathbb{R}^d}{argmax} \sum ||B_i\bar{v}||^2 \quad s.t \enspace||\bar{v}^2||=1
\end{equation*}

where $B_i$ is the block on the diagonal of $M$ for point i. This can be rewritten as $\sum_i \bar{v}^TB_i^TB_i\bar{v} +\lambda(\bar{v}^T\bar{v}-1)$.  To find the maximum, we take the derivative with respect to $\bar{v}$ and set it to zero: $\frac{d}{d\bar{v}} \sum_i \bar{v}^TB_i^TB_i\bar{v} +\lambda(\bar{v}^T\bar{v}-1) = \sum 2B_i^TB_i\bar{v}-\lambda2\bar{v}=0$.  The best perturbation vector is the eigenvector of the matrix $\sum_i B_i^TB_i$ with the largest eigenvalue.
This gives us a single perturbation, $\bar{v}$, that when applied to all points maximizes the change in the projection. $\bar{v}$ is constrained to have unit length to prevent the method from choosing an arbitrarily large perturbation.

\subsubsection{Perturb each point individually}\label{sec:searchIndivid} The second method for recovering the best perturbation is to find different perturbations for each point that collectively change the projection the most constrained so that points that are projected to similar places have similar perturbations.  The formulation is as follows:

\begin{equation*}
\underset{\bar{v}\in\mathbb{R}^d}{argmax} \sum_i ||B_i\bar{v}_i||^2-\lambda\sum_i\sum_j||\bar{v}_i-\bar{v}_j||^2S(i,j) \quad s.t \enspace||\bar{v}^2||=1
\end{equation*}

where $B_i$ is the block on the diagonal for point i, $\bar{v}_i$ is the perturbation for the point i, $\lambda$ is a free parameter for how much smoothing we want, and $S(i,j)$ is the similarity between the projection of points i and j, $p_i$ and $p_j$.  This similarity is defined as $S(i,j) = e^{-||p_i-p_j||^2/\sigma^2}$. $\sigma^2$ is a free parameter set by the user that determines how close points have to be in the projection to be considered similar.   We can rewrite the above equation as follows:

\begin{multline*}
\underset{\bar{v}\in\mathbb{R}^d}{argmax} \sum_i \bar{v}_iB_i^TB_i\bar{v}_i-\lambda\sum_i\sum_j\langle \bar{v}_i-\bar{v}_j,\bar{v}_i-\bar{v}_j\rangle S(i,j) \quad \\
s.t \enspace||\bar{v}^2||=1
\end{multline*}

We observe that $\sum_i\sum_j\langle \bar{v}_i-\bar{v}_j,\bar{v}_i-\bar{v}_j\rangle S(i,j)$ takes form similar to a Laplacian matrix, $L_s$ multiplied by the entire perturbation vector $v$ (the concatenation of all of the individual perturbations, $\bar{v}_i$) on both sides: $\bar{v}^TL_s\bar{v}$. $L_s$ differs from a standard Laplacian matrix in that rather than having diagonal values $\sum_{j\neq i}S(i,j)$ and off diagonal values $-S(i,j)$, it has diagonal values $I*\sum_{j\neq i}S(i,j)$ and off diagonal values $I*-S(i,j)$ where $I$ is $d\times d$ identity matrix. 

Furthermore, the equation can be rewritten in terms of the whole matrix, $M$, and the entire perturbation vector, $\bar{v}$ giving us the following equation which incorporates the constraint on the length of $\bar{v}$:

\begin{equation*}
\underset{\bar{v}\in\mathbb{R}^d}{argmax} \enspace \bar{v}^TM^TM\bar{v}-\lambda \bar{v}^TL_s\bar{v}-\lambda_2\bar{v}^T\bar{v}-1
\end{equation*}

Taking the derivative with respect to $\bar{v}$ and setting it to zero, we get 

\begin{equation*}
\bar{v}^T(M^TM-\lambda L_s)-\lambda_2\bar{v} = 0
\end{equation*}

The entire perturbation vector, $\bar{v}$, is the eigenvector of the matrix $M^TM-\lambda L_s$ with the largest eigenvalue.

\paragraph{Choosing $\lambda$ and $\sigma$.} $\sigma$ controls the width of a gaussian centered on each point. Examining the results of the projection itself gives some information about plausible values for $\sigma$. For example, points outside further than three standard deviations from each other are essentially treated independently, but at the same time, we don't want a $\sigma$ that creates a gaussian which covers the entire projection. For choosing $\lambda$, we should be looking at the resulting perturbations.  If a single point is heavily dominating the perturbation (i.e. it moves much more than the rest of the points) then $\lambda$ is likely too small.  In contrast, if all points are perturbed in almost exactly the same way, this is an indication that $\lambda$ may be too large.


\section{Implementation and Experiments~\label{sec:implementation}}
In this section, we discuss the implementation of our techniques along with a suite of experiments designed to explore
the capabilities, performance, and limitations of \oursystem. We will show how \oursystem\ directly addresses the following gaps identified in Sedlmair et al.'s interview study about gaps between theory and practice in dimensionality reduction (DR)~\cite{sedlmair2012dimensionality}. These include the \emph{interpretation gap}: ``what do the results mean?''; \emph{guidance gap}, ``what algorithm to use?'', and the \emph{non-linear unmapping gap}: ``how do projection dimensions relate to input dimensions?''.

Our current prototype for \oursystem\ is implemented in Python and
numpy~\cite{walt2011numpy}. Our t-SNE implementation is closely based
on van der Maaten's Python code~\cite{maaten-tsne-python}, while the
LLE and Isomap implementations are from-scratch. The entire method
takes about 3,500 lines of Python, including implementations of
Marching Squares, the classes for autodiff, and the linear solvers
described below.

\subsection{DimReader}\label{sec:dimReaderExpir}
In the following we look at a well known dataset, the iris dataset, with known perturbations, and simple
projection algorithms, in order to better understand the behavior of
the technique~\cite{kirby2008need}.
\subsubsection{Linear projections}
We start with showing results of linear projections as a basic sanity check on the behavior of \oursystem. Figure~\ref{fig:iris-all} shows a typical example of the axes reconstructed by \oursystem\ when using linear projections. Since linear projections can be exactly represented by a matrix multiplication, the derivatives of input points position with respect to one direction will always be constant vectors. As a result, the reconstructed scalar field is almost (except for the influence of the regularization terms) a linear ramp, and so the contour lines are evenly spaced and parallel, which indicate that changes in the input variable will behave identically across the entire field. Despite their limited power, this property is one of the main advantages of linear projections.

\subsubsection{Isomap}
Isomap was one of the first NDR techniques to recover curved manifolds
well in practice~\cite{tenenbaum2000global}. Isomap builds a weighted
graph which approximates the manifold, where edges have weight equal
to the distance between points, and each point has edges to its $k$ nearest neighbors
($k$ is specified by the user). The global distance between two points is defined to be the
shortest-path metric on the graph. The low-dimensional projection is
constructed from the shortest-path metric using classical
MDS~\cite{borg2005modern}.

We implemented Isomap not only because of its historical significance
and relatively high-quality results, but also because it highlights an
interesting property of automatic differentiation: it works over code
bases that we tend to not think of as differentiable. Specifically, the
operations in Dijkstra's algorithm for shortest paths are all well
defined for dual numbers, and so we naturally can extract the
sensitivity of shortest-path distances with respect to changes in the
input points~\cite{cormen2009introduction}.

%

\paragraph*{Interaction with numerical linear algebra routines}
The final step of Isomap is Classical MDS, and this presents unique
challenges for our autodiff implementation based on operator
overloading. Specifically, Classical MDS requires the computation of
eigenvectors, and since Python libraries for numerical linear algebra
are implemented through high-performance libraries like Lapack, the
operator-overloading functionality is not present. To solve this
issue, we implement the eigenvalue computation through power
iterations~\cite{golub2012matrix}, since matrix-vector multiplication
of dual numbers has efficient dual-number implementations in terms of
matrices of values and $\epsilon$ terms.

\paragraph*{Isomap Experiments}
Because Isomap uses classical MDS (which is essentially a linear projection), we should expect that, to some degree, Isomap would behave much like linear projections. This is indeed the case with simpler datasets, such as the Iris dataset, shown in Figure~\ref{fig:iris-all}. However, there are some interesting differences. Consider the generalized axis for the ``sepal width'' variable which \oursystem\ recovers. Even though the point positions generated by Isomap are quite similar to that of PCA, the sensitivity of the projection differs dramatically from the cluster of Setosa samples to that of Virginica and Versicolor samples. Even more interestingly, it seems that the sensitivity is caused by only some of the Setosa samples. This differentiation is not present in the linear projection, and would not be clear from the Isomap plot alone. In this example, \oursystem\ helps overcome Sedlmair's \emph{interpretation gap} by providing an explanation for why Isomap spread the points in the Setosa cluster (Isomap is sensitive to differences in the Sepal Width in this cluster) that would otherwise be unknown. 

\begin{figure}
\centering
\includegraphics[width=.85\linewidth]{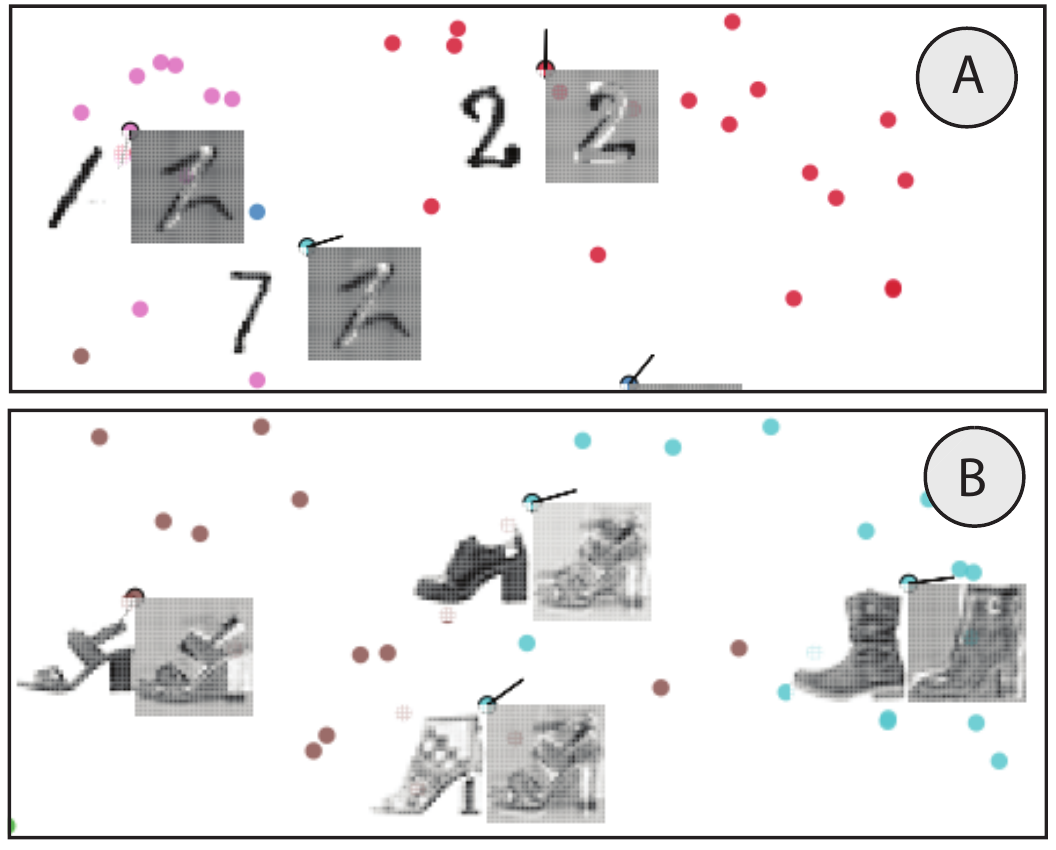}
\caption{\label{fig:pertOverview} An overview of perturbations for points in the(A) MNIST digits and (B) MNIST fashion.  There is structure in the perturbations that our technique discovers.  They often resemble their true digit (or clothing article) but with some variation. Darker areas are perturbed more than lighter areas. }
\end{figure}

\subsubsection{Locally Linear Embedding}
The next algorithm we highlight is Roweis and Saul's Locally Linear
Embedding~\cite{roweis2000nonlinear} (LLE). Like Isomap, LLE uses a
nearest-neighbor graph to recover a global view of the dataset.
LLE computes edge weights for the
nearest neighbor graph, such that each vertex can be best
reconstructed by a linear combination of its neighbors using those
weights. On a second step, the projection coordinates are recovered by
finding positions on the plane that best respect the weights.

\paragraph*{Interaction with numerical linear algebra routines}
Similarly to Isomap, our autodiff implementation of LLE involves a
small degree of adaptation. In the case of Isomap, we required the
computation of the largest eigenvalues of a matrix. In the case of
LLE, we need to compute the \emph{smallest non-zero} eigenvalues. Our
implementation uses \emph{inverse power
  iteration}~\cite{golub2012matrix}.  Inverse power iteration, in
turn, requires a linear system solver, which presents similar issues
for dual number implementations. Our solution is to implement a
black-box linear system solver using conjugate
gradients~\cite{shewchuk1994introduction}.





\paragraph{LLE Experiments} Locally Linear Embedding is a popular method due to its performance~\cite{roweis2000nonlinear}, but is known to produce distorted projections~\cite{donoho2003hessian}. In this section, we illustrate how \oursystem\ might help pinpoint such problems.  Consider the projection of the iris dataset in Figure~\ref{fig:iris-all}. Note that neither of the recovered axes quite cross the projection perpendicularly on the left side of the arc (the Versicolor and Virginica cluster): no direction of perturbation on the input moves the points diagonally along that cluster. This suggests that the shape of the cluster is an artifact of the projection method. Compare this with the Isomap projection in Figure~\ref{fig:iris-all}: Isomap has perturbations which cross each of the clusters perpendicularly (Sepal Width for Setosa and Sepal Length for Versicolor and Virginica). Thus, Isomap is more faithful to the underlying data than LLE.
This is evidence that \oursystem\ helps bridge Sedlmair et. al's \emph{guidance gap}~\cite{sedlmair2012dimensionality}, giving an indication for which NDR algorithm performs better for this data.



\subsubsection{t-SNE}
t-SNE is among the most powerful techniques
for dimensionality reduction, and also one of the hardest to interpret
appropriately~\cite{maaten2008visualizing, wattenberg2016how}. As
such, it is a natural target for \oursystem. In addition, t-SNE is
significantly different from Isomap and LLE in both formulation and
implementation. This provides us with an opportunity to explore
practical issues of using \oursystem\ to explain its results.

We highlight two separate issues to discuss: the presence of
multiple local minima, and its formulation in terms of the
gradient of an energy function. While the first issue presents challenges
for implementations that depend on repeated executions, the second
issue allows us to achieve a significant speedup.

\paragraph*{Multiple minima} The energy function that t-SNE minimizes
has more than one local minimum. This means that any source of
randomness in the implementation will cause multiple runs to possibly
diverge, presenting a challenge for our approach. Most implementations
of t-SNE require an initial guess for the projection, and we take
central advantage of this. Specifically, in our first execution of t-SNE
we use a random initial guess and regular floating-point numbers to 
calculate a local minimum that is then used as the initial guess for subsequent runs.
In the initial run we also capture variables that serve as parameters for
subsequent runs to ensure that they reach the same local minimum. 

\paragraph*{Gradient descent}
t-SNE is implemented as an explicit gradient descent formulation
through an additive update of the parameters. Specifically,
the main loop of t-SNE is roughly as follows:

\begin{lstlisting}[language=Python]
pos = initial_guess
g = gradient(energy(pos), pos)
while mag(g) > epsilon:
    pos = pos - rate * g
    g = gradient(energy(pos), pos)
%\end{minted}
\end{lstlisting}

As a result, when the loop exits, we know that the gradient of the
energy with respect to the position will be close to zero. This means
that to recover any one perturbation of the t-SNE formulation with
respect to an input point, all that is required is to run one single
iteration of t-SNE with dual numbers. By providing the dual-number
implementation the result of the execution of the floating-point
implementation (as explained in the previous paragraph), the loop will
execute at most once before exiting -- in fact, in order for the
sensitivity of the positions with respect to the input to be recorded
in the   \lstinline[language=Python]{pos} variable, we must force the loop to
execute at least once. Still, since t-SNE typically executes between
100 and 1000 iterations in this loop, this simple optimization
achieves a significant speedup.

\paragraph{t-SNE Experiments}
T-SNE is often considered to be the state of the art in NDR methods, but one of the main objections to its use in practice is the opaque nature of its optimization criteria~\cite{wattenberg2016how}.
It is unclear how effectively the projection recovers high dimensional information. Consider the t-SNE axes in Figure~\ref{fig:iris-all}.
t-SNE is detecting variation in the petal length and petal width of the Virginica and Versicolor cluster and subsequently spreading the cluster based on these dimensions.
This helps explain how petal length behaves in the projection, providing evidence that \oursystem\ helps bridge the \emph{non-linear unmapping gap}~\cite{sedlmair2012dimensionality}.

\subsection{Discovering Perturbations}\label{sec:discovPertExpir}
The implementation of the equations in ~\ref{sec:searchingPerts} for discovering perturbations is straightforward as long as the machine has sufficient memory to hold the expanded laplacian matrix, $L_s$.
This matrix becomes very large for very high-dimensional data and thus requires significant memory. To solve this problem, we were again able to exploit the block structure of the tangent map as well as the structure of the Laplacian matrix: the diagonal values are $\sum_{j\neq i} S_{i,j}$ and the off diagonal values are $-S_{i,j}$.  We implemented a version of power iteration that does not require access to the matrix $M^TM-\lambda L_s$ but instead requires a function that, when given a vector $v$, returns $(M^TM-\lambda L_s)v$. The multiplication function calculates elements of the output vector individually and thus does not require the entire $L_s$ matrix. 
%

Using this method, we uncovered perturbations for several datasets projected with t-SNE. In the following experiments, we use the method in Section~\ref{sec:searchIndivid} to find individual perturbations for each point.

\subsubsection{Iris}
We first look at the best perturbation for the Iris dataset. Figure~\ref{fig:IrisPert} shows the DimReader plot for this perturbation as well as a plot for each dimension that shows how much that dimension was perturbed in each point through the color (the darker purple a point is, the more it was perturbed). The DimReader plot shows that the best perturbation only perturbs points in the Setosa (red) cluster.  In the individual dimension plots, the Setosa cluster is perturbed primarily in the Sepal length and Sepal width dimensions which tells us that in this projection, points in the Setosa cluster are sensitive to changes in the Sepal dimensions.  Comparing~\ref{fig:IrisPert} to the t-SNE plots in Figure \ref{fig:iris-all}, the movement of the Setosa cluster with the discovered perturbation is similar to the movement when the sepal width or sepal length is perturbed.

\subsubsection{MNIST Digits}
Figure~\ref{fig:pertOverview} (A) shows an sample of discovered perturbations in the projection. The perturbation images often resemble a variation of their corresponding digit or a nearby digit (due to the constraints defined in Section ~\ref{sec:searchIndivid}).  These perturbations show us that t-SNE is capturing meaningful information about the dataset.  In Figure~\ref{fig:pertOverview} (A), the perturbation that moves the "seven" the most turns the "seven" into a ``two'' and moves it toward the cluster of ``two''s. Thus, \oursystem,  is showing evidence that t-SNE is capturing information about what constitutes a ``two'' and is using that information to separate out the two's into their own (imperfect) cluster.

\begin{figure}
\centering
\includegraphics[width=.78\linewidth]{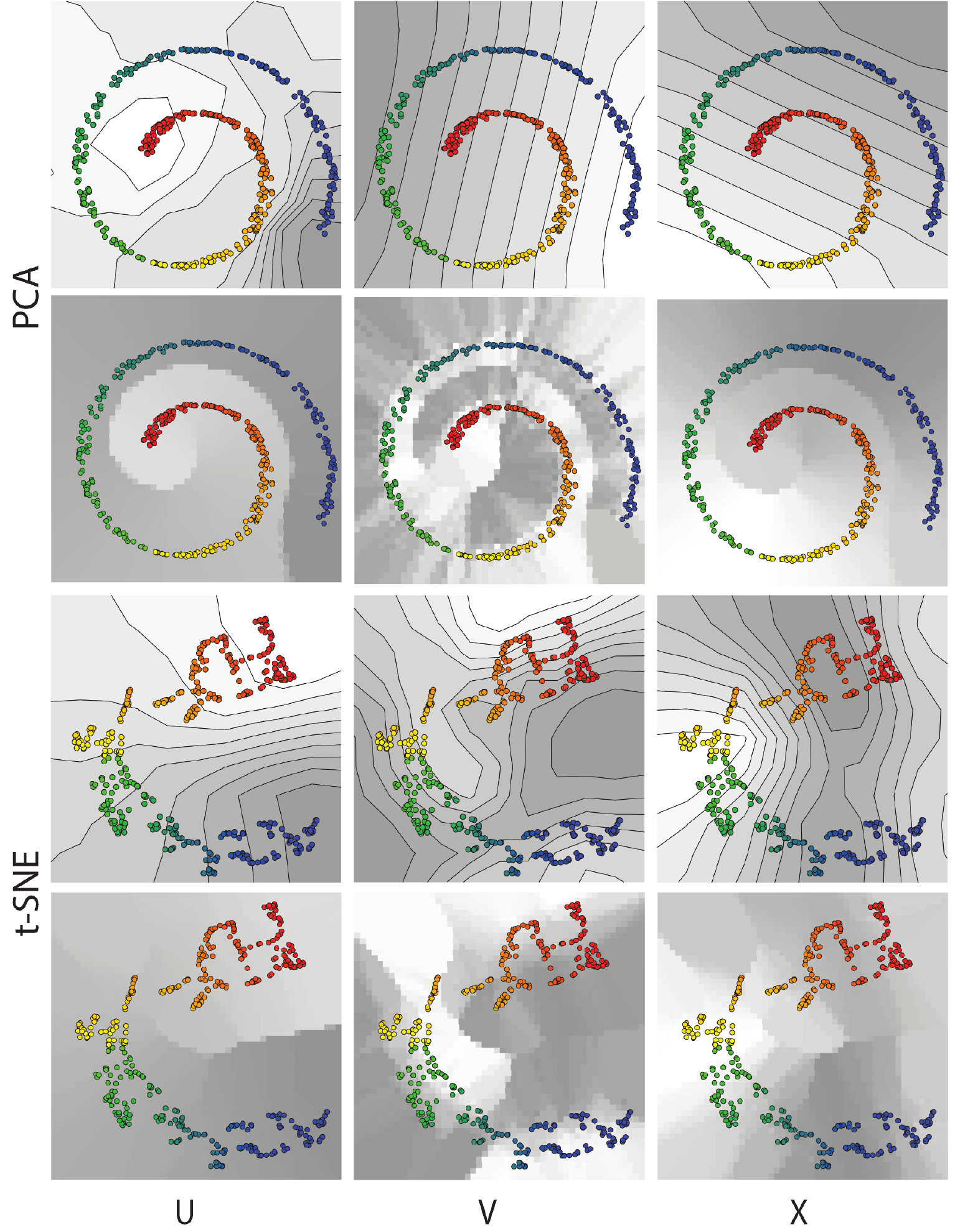}
\caption{\label{fig:swissProbeComp} DimReader axes and value heatmaps for the u,v, and x dimensions of the swiss roll.  A discussion of these plots is in Section~\ref{sec:synthetic}.}
\end{figure}

\subsubsection{MNIST Fashion}
The MNIST Fashion dataset is similar to the digits dataset in that each point represents a $28\times28$ pixel image and there are 10 different classes of images (articles of clothing) but is more complicated than the digits dataset.  A t-SNE plot with selected perturbations found by our technique is shown in Figure~\ref{fig:pertOverview} (B).  Just as in the digits perturbations, there is structure in these perturbations. In the example in Figure 1 (C), our technique is finding perturbations that capture information about how t-SNE is projecting the data.  \oursystem tells us, that for the heel in the middle, the perturbation that moves this point the most, changes it from a heel into a flat shoe.  This also shows us that t-SNE understand the difference between flat shoes and heels and is able to separate them.

\new{\subsection{Synthetic Examples\label{sec:synthetic}}
In this section we will look at the DimReader plot with two synthetic examples, the swiss roll and the interlocked rings, and compare them to the value heatmaps for each dimension from Stahnke et. al's Probing Projections.

\subsubsection{Swiss Roll} The swiss roll dataset is calculated from the equations $x = u \cos(u)$, $y = u \sin(u)$,  $z=v$ where $\frac{3\pi}{2} \leq u\leq \frac{9\pi}{2}$ and $0\leq v\leq15$.  Figure ~\ref{fig:swissProbeComp} shows perturbations of the U,V, and X dimensions.  

\begin{figure}
\centering
\includegraphics[width=.9\linewidth]{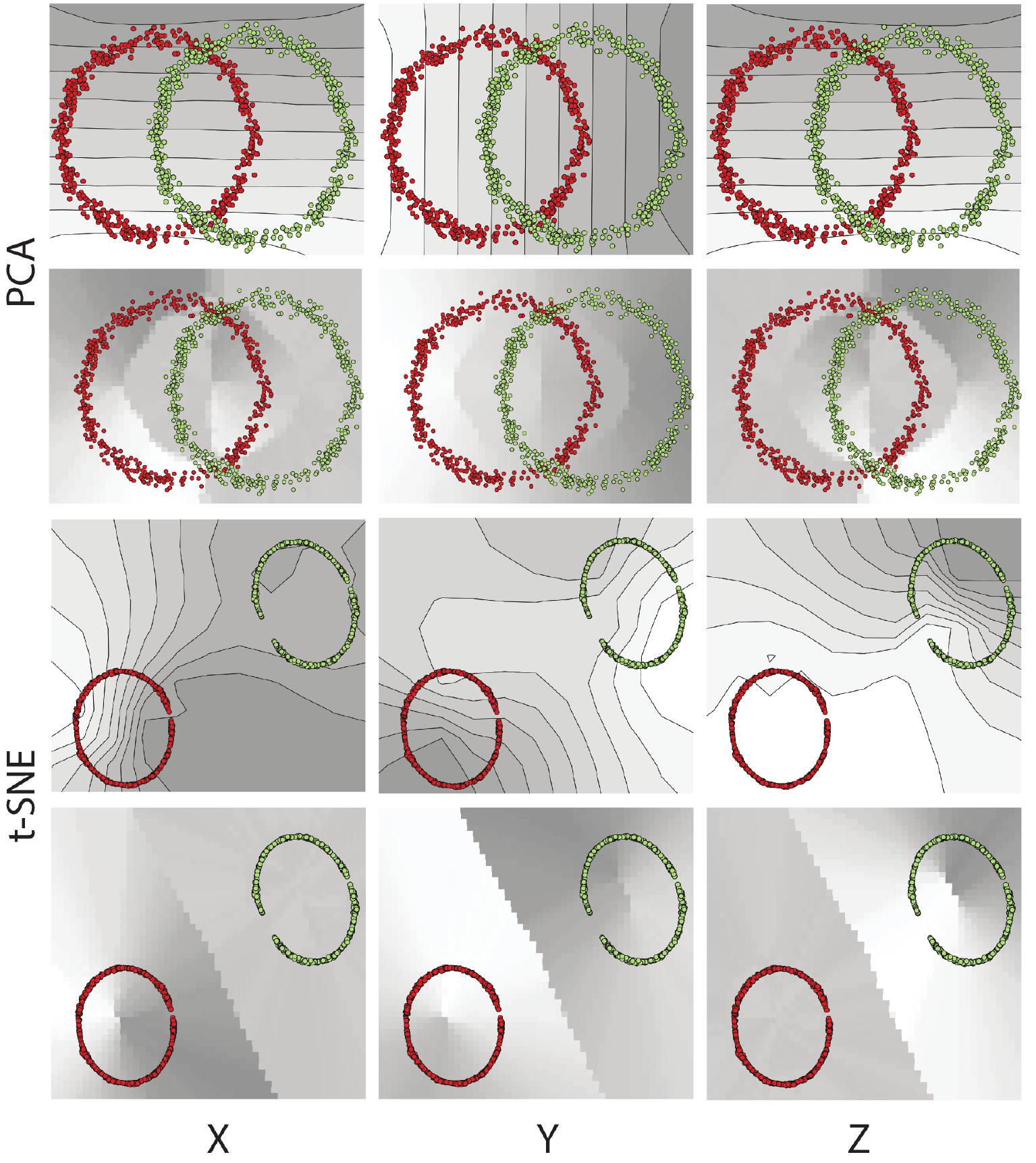}
\caption{\label{fig:ringsProbeComp} DimReader axes and value heatmaps for the x, y, and z dimension of the interlocked rings. A discussion of these plots is in Section~\ref{sec:synthetic}.}
\end{figure}

\paragraph{PCA}
In the $v$ dimension, the DimReader plot shows that increasing the $v$ in the original data moves the projected points to the left.  In comparison, the value heatmap for the $v$ dimension is difficult to read due to the high variance in $v$ between neighboring points.  This highlights a fundamental difference between DimReader and value heatmaps: DimReader is showing what the \emph{projection} is doing while value heatmaps show the values of a dimension based on the placement of points. If we created a projection that simply mapped each point to its PCA coordinates through a table lookup, the value heatmaps would not change whereas the DimReader plots would show nothing because changing any of the dimensions would not change the projection. \rf{not sure about the last part without an actual example, it's true but we don't have supporting example}

The DimReader plot for the $u$ dimension shows that changing the $u$ dimension would move the point along the spiral, from red to blue.  The isolines, however, are irregular from green to orange.  These irregularities could be due to the resolution of the grid or the regularization.  One direction for future work is to automatically determine the appropriate grid resolution and regularization for a projection.

\paragraph{t-SNE}
In the DimReader plot for t-SNE, the $u$ dimension behaves exactly as we would expect, increasing as we move from red to blue.  The spacing of the lines around the curve, specifically how the curves are wider on the outside than on the inside, indicates the bending does not reflect the underlying data but rather is caused by the projection. 

In the DimReader plot for $v$, t-SNE has flipped the green and yellow segment (the points move to the bottom left rather than the upper right). This appears less clear from the value heatmap alone. 

In the plots for $x$, the highest values in the heatmap do not match the area in the DimReader plot where the biggest change occurs.

\subsubsection{Interlocked Rings}
The DimReader plots and value heatmaps generated for the interlocked rings dataset are shown in Figure~\ref{fig:ringsProbeComp}.

\paragraph{PCA}
The PCA plots from DimReader show that changing X or Z would move the points upward and changing the Y dimension move points to the right.  Again, the plots here highlight the difference between our technique and the value heatmaps: the heatmaps of x and z tell us that the x and z values are only changing over one ring each whereas our technique is showing that PCA will move the rings vertically if the x or z dimension is changed.  DimReader shows what the projection does, whereas heatmaps shows where the values go.

\paragraph{t-SNE} T-SNE has separated the two rings well.  In the DimReader plot of the X dimension, the red ring will move from left to right when the X dimension is changed.  Comparing the X dimension to the Y dimension, for the red ring, the two axes are nearly perpendicular to one another.  This suggests that t-SNE is primarily using these two dimensions for projecting the red ring. Furthermore, in the red ring the X dimension changes much quicker than the Y dimension (the lines are closer together) which indicates that t-SNE is distorting the shape of the red ring.  Similar observations can be made about the green ring with the Y and Z dimensions.  Again, the Y and Z are nearly perpendicular in the green ring.  In the Y dimension, the axes change their behavior when they reach the gap in the green ring. Points in this region move more slowly when changed which in turn tells us that this is likely a tear in the ring caused by t-SNE that does not reflect the structure of the underlying data.  It is not as clear from the value heatmaps that the gap is a tear in the data caused by the projection.
}



\begin{figure}
  \centering
  \includegraphics[width=\linewidth]{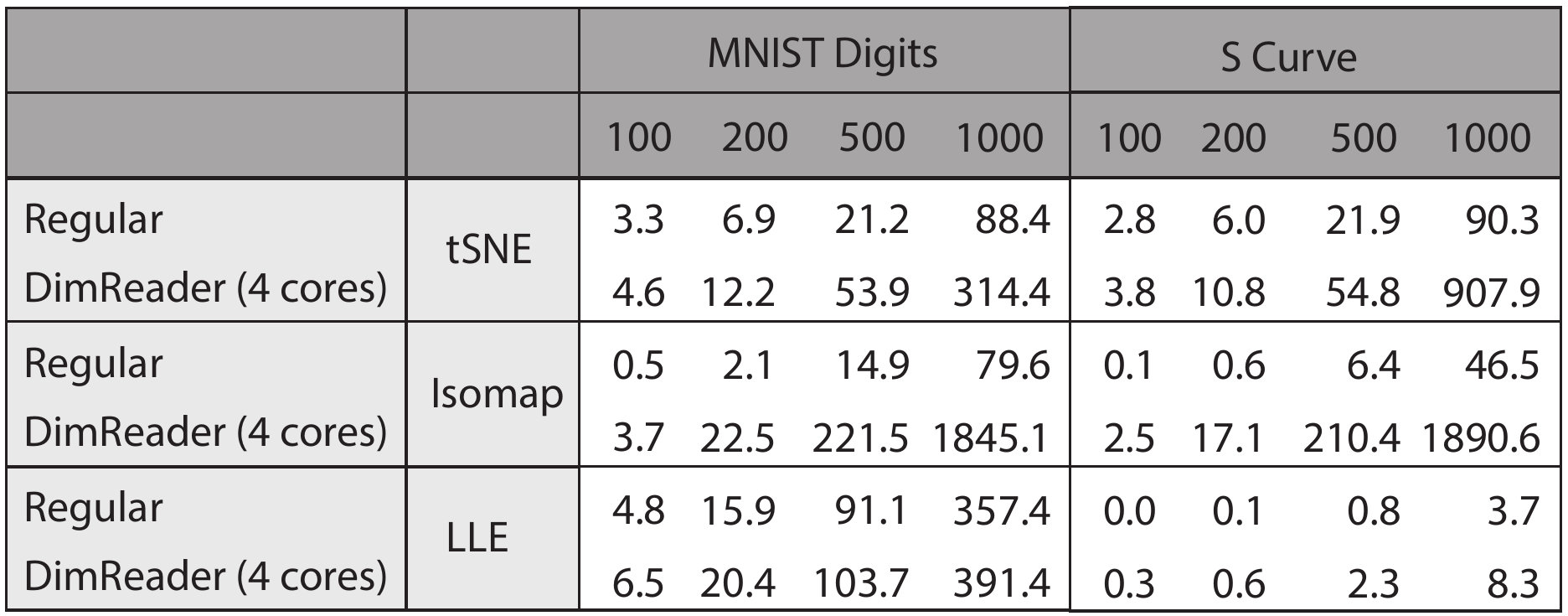}
  \caption{\label{fig:performance} Performance figures for the MNIST digits
    dataset and the S-Curve dataset, for progressively larger samples
    and three different NDR methods. All figures are reported in
    seconds.}
\end{figure}

\subsection{Performance\label{sec:performance}}
\subsubsection{Known Perturbations}
In this section, we report performance figures for the prototype
implementation of \oursystem. Although we were reasonably careful with
algorithmic and high-level design decisions that impact performance,
we did not make a significant effort to make \oursystem\ fast. We
expect carefully-implemented versions of our proposal in
high-performance languages such as C++ or Java to be significantly
faster, possibly by an order of magnitude (typically the
performance difference between Python and aggressively optimized,
compiled languages).

A table showcasing typical results is included in
Figure~\ref{fig:performance}. The performance of \oursystem\ for a
given NDR method is dependent on two main factors: the number of input
points and the overhead incurred by dual numbers. We need to execute a
number of repeated runs proportional to the base-2 logarithm of the
number of input points, and that is essentially unavoidable. We note
that for the case of LLE and t-SNE, the optimizations we described in
the previous section make the execution of the dual-number version of
the projection much faster than that of the regular numbers. As a
result, \oursystem\ can extract axes with a relatively small
performance overhead.

For cases such as Isomap, on the other hand, where we performed no
such optimizations, the performance of our method suffers a bit.  We
argue that this is an acceptable tradeoff: \oursystem\ still works in
an acceptable amount of time in the general case, but more careful
implementations can be significantly more efficient.

\subsubsection{Discovering Perturbations}
The most expensive part of searching for a perturbation is calculating the tangent map.  The tangent map is $n*d\times n*d$ and requires $d$ executions of \oursystem\ to build it.  For datasets with a large number of data points and dimensions, this quickly becomes slow.  Once we have the matrix, the performance for finding the best perturbation largely depends on whether or not we can the expanded Laplacian matrix, $L_s$, (described in Section~\ref{sec:discovPertExpir}) in memory.  If we can't and have to use power iteration, the performance depends on the speed of our multiplication function as well as the amount of time it takes power iteration to converge.  We did not make a significant effort to increase the performance for calculating the matrix or searching for perturbations; this remains for future work.  


\section{Discussion~\label{sec:discussion}}


\paragraph*{Can we trust DimReader plots?}
While we have shown that DimReader can help determining how
NDR plots can be trusted, a natural question to ask is: can the
DimReader plots themselves be trusted? One natural scenario in which
this comes up is when perturbation vectors of nearby projections
disagree with one another. It's always possible to show the
vectors themselves as a diagnostic of the quality of the reconstructed
axis lines, but a proper, user-centric evaluation of the settings in which
\oursystem's axes are more informative than naked NDR plots is clearly
necessary, and will be the subject of future work.

\paragraph*{Inverse readings}
\oursystem\ enables interpretation of forward transformations: given a
perturbation of an input and a visualization, \oursystem\ provides an
answer. But a different natural reading is the
\emph{inverse}: given a projected point and a direction of movement in
the projection, what changes in the data could generate such movement?
In principle, the derivative information obtained by autodiff also
captures this inverse relationship~\cite{carmo1994differential}, but
the fact that we are dealing with \emph{projections} makes
the problem fundamentally harder. A full investigation is beyond the
scope of this work.

\paragraph*{More algorithms, better infrastructure}
While \oursystem\ shows that it is possible to adapt a large number of
existing NDR methods to run within an autodiff framework, one
goal is to provide \oursystem\ axes to as much existing
visualization infrastructure as practically possible. In such
scenarios, reducing the implementation effort even further would be
desirable. The majority of our difficulties porting algorithms to
automatic differentiation arose due to difficulties in evaluating
derivatives of linear-algebraic concepts, such as solutions of a
linear system and eigenvectors. Some of these have explicit
formulas~\cite{petersen2008matrix}, but incorporating them in an
autodiff system effectively and efficiently is a
fundamental challenge beyond the scope of our work.  We note, in
addition, that our choice of automatic differentiation is not strictly
necessary. Other methods exist to evaluate function derivatives,
including manual derivation of the expressions.  When using
\oursystem\ with a specific NDR method, these alternatives
might be more attractive. This might be particularly true whenever
approximations of the derivative can be computed more efficiently than
autodiff.

\section{Conclusion}
In this paper, we identified \emph{infinitesimal perturbations} as a tool to enable interpretation of NDR plots, and presented \oursystem, a technique that produces generalized axes for studying such perturbations. While much work remains to be done, \oursystem\ strikes a favorable balance between generality and power, highlighting strengths and weaknesses of a variety of NDR methods, and providing a novel perspective into what NDR methods are actually visualizing.

\paragraph{Acknowledgements} We acknowledge fruitful discussions of dimensionality reduction with Sean Stephens, Luiz Gustavo Nonato, and Joshua Levine on the energy formulation of t-SNE. This work has been partially supported by the NSF under the TRIPODS program, award number CCF-1740858, and IIS-1513651.

\bibliographystyle{IEEEbib}

\end{document}